\documentclass [a4paper,12pt]{article}
\usepackage{amsmath}
\usepackage{amssymb}
\usepackage{graphics}
\usepackage{picins}
\renewcommand{\thetable}{\Roman{table}}

\newcommand{\CO}{CO$_2$}
\newcommand{\COc}{[CO$_2$]}

\catcode`\_=\active
\def_{\sb\mathrm}

\DeclareSymbolFont{EUr}{U}{eur}{m}{n}
\DeclareSymbolFont{EUb}{U}{eur}{b}{n}
\DeclareMathSymbol{\upGamma}{\mathord}{EUr}{"00}
\DeclareMathSymbol{\upDelta}{\mathord}{EUr}{"01}
\DeclareMathSymbol{\upTheta}{\mathord}{EUr}{"02}
\DeclareMathSymbol{\upLambda}{\mathord}{EUr}{"03}
\DeclareMathSymbol{\upXi}{\mathord}{EUr}{"04}
\DeclareMathSymbol{\upPi}{\mathord}{EUr}{"05}
\DeclareMathSymbol{\upSigma}{\mathord}{EUr}{"06}
\DeclareMathSymbol{\upUpsilon}{\mathord}{EUr}{"07}
\DeclareMathSymbol{\upPhi}{\mathord}{EUr}{"08}
\DeclareMathSymbol{\upPsi}{\mathord}{EUr}{"09}
\DeclareMathSymbol{\upOmega}{\mathord}{EUr}{"0A}
\DeclareMathSymbol{\upalpha}{\mathord}{EUr}{"0B}
\DeclareMathSymbol{\upbeta}{\mathord}{EUr}{"0C}
\DeclareMathSymbol{\upgamma}{\mathord}{EUr}{"0D}
\DeclareMathSymbol{\updelta}{\mathord}{EUr}{"0E}
\DeclareMathSymbol{\upepsilon}{\mathord}{EUr}{"0F}
\DeclareMathSymbol{\upzeta}{\mathord}{EUr}{"10}
\DeclareMathSymbol{\upeta}{\mathord}{EUr}{"11}
\DeclareMathSymbol{\uptheta}{\mathord}{EUr}{"12}
\DeclareMathSymbol{\upiota}{\mathord}{EUr}{"13}
\DeclareMathSymbol{\upkappa}{\mathord}{EUr}{"14}
\DeclareMathSymbol{\uplambda}{\mathord}{EUr}{"15}
\DeclareMathSymbol{\upmu}{\mathord}{EUr}{"16}
\DeclareMathSymbol{\upnu}{\mathord}{EUr}{"17}
\DeclareMathSymbol{\upxi}{\mathord}{EUr}{"18}
\DeclareMathSymbol{\uppi}{\mathord}{EUr}{"19}
\DeclareMathSymbol{\uprho}{\mathord}{EUr}{"1A}
\DeclareMathSymbol{\upsigma}{\mathord}{EUr}{"1B}
\DeclareMathSymbol{\uptau}{\mathord}{EUr}{"1C}
\DeclareMathSymbol{\upupsilon}{\mathord}{EUr}{"1D}
\DeclareMathSymbol{\upphi}{\mathord}{EUr}{"1E}
\DeclareMathSymbol{\upchi}{\mathord}{EUr}{"1F}
\DeclareMathSymbol{\uppsi}{\mathord}{EUr}{"20}
\DeclareMathSymbol{\upomega}{\mathord}{EUr}{"21}
\DeclareMathSymbol{\upvarepsilon}{\mathord}{EUr}{"22}
\DeclareMathSymbol{\upvartheta}{\mathord}{EUr}{"23}
\DeclareMathSymbol{\upvaromega}{\mathord}{EUr}{"24}
\DeclareMathSymbol{\upvarphi}{\mathord}{EUr}{"27}
\renewcommand{\Gamma}{\upGamma}
\renewcommand{\Delta}{\upDelta}
\renewcommand{\Theta}{\upTheta}
\renewcommand{\Lambda}{\upLambda}
\renewcommand{\Xi}{\upXi}
\renewcommand{\Pi}{\upPi}
\renewcommand{\Sigma}{\upSigma}
\renewcommand{\Upsilon}{\upUpsilon}
\renewcommand{\Phi}{\upPhi}
\renewcommand{\Psi}{\upPsi}
\renewcommand{\Omega}{\upOmega}
\renewcommand{\alpha}{\upalpha}
\renewcommand{\beta}{\upbeta}
\renewcommand{\gamma}{\upgamma}
\renewcommand{\delta}{\updelta}
\renewcommand{\epsilon}{\upepsilon}
\renewcommand{\zeta}{\upzeta}
\renewcommand{\eta}{\upeta}
\renewcommand{\theta}{\uptheta}
\renewcommand{\iota}{\upiota}
\renewcommand{\kappa}{\upkappa}
\renewcommand{\lambda}{\uplambda}
\renewcommand{\mu}{\upmu}
\renewcommand{\nu}{\upnu}
\renewcommand{\xi}{\upxi}
\renewcommand{\pi}{\uppi}
\renewcommand{\rho}{\uprho}
\renewcommand{\sigma}{\upsigma}
\renewcommand{\tau}{\uptau}
\renewcommand{\upsilon}{\upupsilon}
\renewcommand{\phi}{\upphi}
\renewcommand{\chi}{\upchi}
\renewcommand{\psi}{\uppsi}
\renewcommand{\omega}{\upomega}
\renewcommand{\varepsilon}{\upvarepsilon}
\renewcommand{\vartheta}{\upvartheta}

\renewcommand{\varphi}{\upvarphi}

\usepackage{sidecap}
\usepackage[sf]{caption}

\renewcommand{\figurename}{Fig.}
\renewcommand{\tablename}{Table}
\renewcommand{\thetable}{\Roman{table}}
\renewcommand{\thefigure}{\arabic{figure}}
\makeatletter
\renewcommand{\fnum@figure}{\sffamily\textbf{\figurename~\thefigure}}
\renewcommand{\fnum@table}
{\sffamily\textbf{\tablename~\thetable}}

\begin{document}

\title{\LARGE \bf Phase relation between global temperature and atmospheric carbon dioxide}

\author{Peter Stallinga$^1$, Igor Khmelinskii$^2$\\
1: University of The Algarve\\
Faculty of Science and Technology\\
Center for Electronics Optoelectronics
and Telecommunications\\
e-mail: pjotr@ualg.pt\\
2: University of The Algarve\\
Faculty of Science and Technology\\
Centro de Investiga\c{c}\~{a}o em Qu\'{i}mica do Algarve\\
e-mail: ikhmelin@ualg.pt
}
\maketitle

\renewcommand{\baselinestretch}{2}
\normalsize

\begin{abstract}

The primary ingredient of Anthropogenic Global Warming hypothesis is the assumption that atmospheric carbon dioxide variations are the cause for temperature variations. In this paper we discuss this assumption and analyze it on basis of bi-centenary measurements and using a relaxation model which causes phase shifts and delays.

Note: This paper was (and is) submitted to various journals and received no scientific criticism whatsoever. It is being rejected principally on reasons of format or for having an alleged political agenda. Scientific comments can be sent to our contact addresses mentioned above.

\end{abstract}
Key--Words: Global Warming, cause and effect, relaxation model
\newpage

\section{Introduction}

Currently, one of the biggest worries of our society is the future of the climate. Common belief is that our planet is heating up at an accelerated rate, caused by the rapid increase in carbon dioxide (\CO) concentration in the atmosphere, henceforth called \COc. This increased carbon dioxide finds its origin in human activity; humans burn fossil fuels, thereby injecting large quantities of carbon into the troposphere by converting it into \CO. The {\CO} contributes to the greenhouse effect of our atmosphere and it is believed that the anthropogenic {\CO} will heat by up the planet by up to six degrees during this century (page 45 of IPCC 2007 report\cite{ipcc2007}). Here we will analyze these ideas and come up with some remarkable conclusions. For that, while the subject is the atmosphere, we do not have to go into much detail of atmospheric science. There are (nearly philosophical) observations one can make about climate systems, even without going into technical details. They are in the realm of signal processing and feedback theory.

The model of Anthropogenic Global Warming (AGW) stands or falls with the idea that temperature is strongly correlated to {\COc} by the so-called greenhouse effect. Serious doubt is immediately found by anybody analyzing the data. The contribution of {\CO} to the greenhouse effect can easily be estimated to be about 3.612\%\cite{montehieb}. The total greenhouse effect is also well known; without our atmosphere our planet would be 32 degrees colder. This makes the {\CO} greenhouse effect only 1 kelvin in a simple analysis. We arrive at a similar value if we use statistics and do a linear regression on contemporary {\COc} and temperature data, the maximum of the effect we can thus expect in a linear model when doubling the concentration artificially by burning up fossil fuels\cite{stalliban}. This is far below the Global Warming models even if we were to use a linear model. It is however, unlikely that the effects are linear. The system is more likely to be sublinear. That is because the greenhouse effect is governed by absorption of light which unavoidably follows the Beer-Lambert Law: the absorption is highly sublinear; twice as much {\CO} will not cause twice as much absorption. The classical Arrhenius' Greenhouse law states that the forcing is logarithmic.

Yet, later models incorporating non-linear positive-feedback effects as proposed by many climate scientists do predict a super-linear behavior and come up with an estimate of between 1.1 and 6.4 degrees heating for the next century as caused by our carbon dioxide injection into the atmosphere\cite{ipcc2007}. The positive feedback can come from secondary effects such as an increase in water in the atmosphere, a strong greenhouse agent, or a \CO-degassing of ground in the permafrost regions when these thaw.

Climate scientists are basing these conclusions mainly on research of the so-called finite-elements type, dividing the system in cells that interact, the same way the weather is studied. Such systems are complicated, but by tuning the processes and parameters that are part of the simulations they manage to explain the actual climate data to an impressive accuracy, as evidenced by the quality of pictures presented in the official climate reports, see for example the IPCC 2007 report where simulation and reality are as good as indistinguishable\cite{ipcc2007}, and, moreover, alarmingly, they conclude that the recent rise in temperature can only be attributed to {\CO}.

But, from a philosophical point of view, the fact that the past was explained very accurately does not guarantee the same quality for the prediction of the future. The climate system is chaotic. Small deviations in parameters and initial conditions or assumptions made in the simulations can cause huge changes in the outcome. This is easily explained in an example from electronics. If we have a chaotic circuit with, for instance, critical feedback, we can go to our SPICE or Cadence simulator and find the parameters of our components that exactly explain the behavior of our circuit. So far so good. The problem is that if we now go back and switch on the same circuit, we will get a different result. (Just take an operational amplifier with 100\% positive feedback, it can saturate at the output at the positive as easily as the negative supply voltage, either one can be simulated). An additional problem is that even the parameters themselves are not constant and seem to change without any apparent reason, for instance the El Ni\~{n}o phenomena in the climate. This is one of the reasons electronic engineers talk about 'phase margins', the zone in Nyquist plots, real vs.\ imaginary parts of gain, that should be avoided because the circuit will become unpredictable even if it is perfectly simulatable.

In fact, recent temperature data fall way out of the prediction margins of earlier models. In view of the discussion above, this does not come as a surprise. Where extrapolation from the 2007 IPCC report predicted 2011 to be a year with an anomaly of close to one degree (0.95 $\rm ^o$C is our personal estimate based on Fig.\ 2.5 of the IPCC 2007 Report), in reality the anomaly is closer to zero. Since 1998, the hottest year in recent history, the planet has actually been cooling, something that was not foreseen by the predictions of 2007 where a continuing exponential increase in temperature was forecasted by the then generally accepted model. The scientific community is now going back to their drawing boards and fine-tunes its models to new perfection and manages to simulate the new data as well. This is a Bayesian way of doing science and is significantly less reliable. The correctness of this statement is evidenced by the fact that there now apparently exist many models that explain the data up to a certain point in time; every correction of the model that is still consistent with earlier data proves this. Apparently, there are a manifold of models that can explain certain data quite satisfactorily (but that diverge for future predictions). In view of this, one should be reluctant in making strong claims about the correctness of the latest model.

Just like in the weather, where the same simulation-evaluation techniques are used, we can only hope to get the predictions reasonably under control after thousands of iterations between predictions and reality. Each iteration takes about the amount of time as the prediction span -- one week with the weather, 30 years with the climate. Honestly speaking, before we get it right, it'll take at least some hundreds of centuries if we uniquely use the approach of finite-elements calculations on supercomputers. In the meantime, we should not see any climate models as proven indisputable facts. A skeptic approach to any scientific model is not an illness, it is an essential ingredient in science. Theories are correct until proven wrong. Ideas that stand up to scrutiny are more likely to be correct than ideas one is supposed to not question. 

Still, undeniably, a strong correlation is found between the {\CO} concentrations and the temperatures as measured by gas-analysis in drillings in ice shelves, see for example the data of the PANGAEA project indicating that one is the function of the other for the past hundreds of thousands of years\cite{pangaea}. That is a very strong point.

\begin{figure}
 \centering
 \scalebox{0.6}{\includegraphics{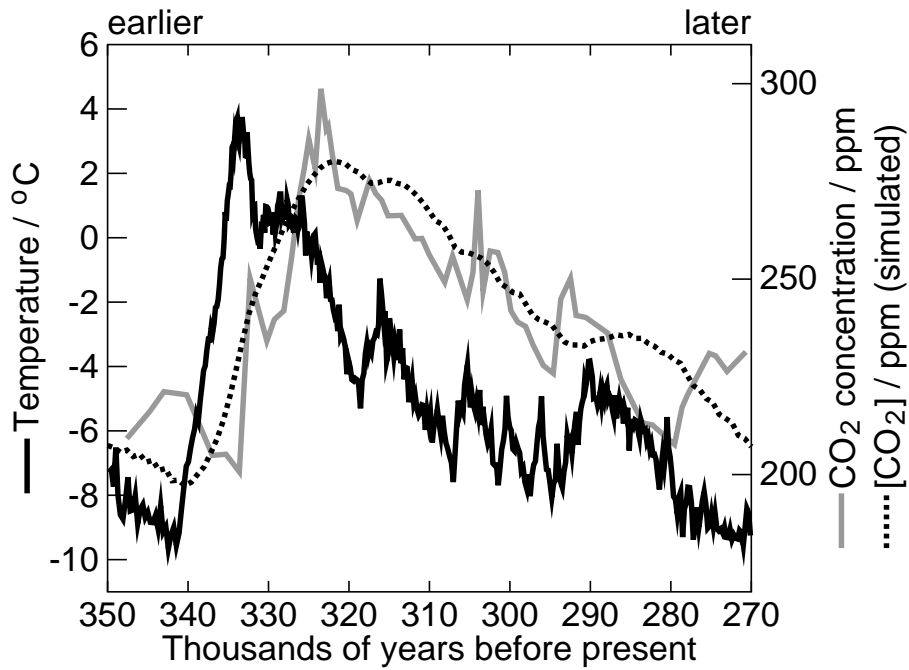}}\\
 \caption{\label{fig:AlGore}
Detail of data of ice shelf drilling correlating the {\CO} concentration and temperature. It is obvious that {\CO} lags behind the temperature. This is consistently the case. A simulation is shown (dashed line) of a convolution of the temperature with a delay of 15 thousand years}
\end{figure}

However, proving only statistical correlation, it is not clear from these data which one comes first. Are temperature variations the result of {\COc} variations, or vice versa? While the data are consistent with the model of AGW they cannot serve as proof of these models. In fact, upon closer scrutiny, the temperature always seems to be \textit{ahead} of {\CO} variations. See Figure \ref{fig:AlGore}, where a detail of the temperature and {\COc} history as measured by ice-trapped gases is plotted, picturing the most blatant example of this effect. A simulation (dashed line) is also shown with an exponential-decay convolution of 15 kyr, quite adequately reproducing the results. Inderm\"{u}hle and coworkers\cite{indermuehle} made a statistical analysis and find a value of 900 yr for the delay and note that "This value is roughly in agreement with findings by Fischer et al.\ who reported a time lag of {\CO} to the Vostok temperature of (600 $\pm$ 400) yr during early deglacial changes in the last 3 transitions glacial-interglacial"\cite{fischer}. This is inexplicable in the framework of Global Warming models and we honestly start having some legitimate doubts.

The apparent time lag may possibly be due to a calibration problem of the measurements, and indeed corrections have been made to the data since then, to make {\COc} variations and temperature variations coincide. While these corrections are the result of circular reasoning, where the magnitude is found by modeling the behavior of ice based on climate models and the climate models based on the ice behavior, these corrections are not even sufficient to remove our doubts. If the correlations are true and we continue to claim that temperature variations are the result of {\COc} variations, something is still not correct. The Vostok data of Figure \ref{fig:AlGore} show a sensitivity of 10 degrees for 50 ppm {\COc}. Contemporary {\COc} are of the order of 80 ppm rise from the preindustrial value. We are thus in for a 16-degree temperature-rise. The fact that we did not reach that level means that either {\CO} is not climate forcing, or that there is a delay between {\COc} variations (cause) and temperature variations (effect). To get a rough idea of the magnitude of this delay, in 25 years, only 2.5\% (0.4 of 16 degrees) of this rise occurred. The relaxation time is thus (25 years)/$\ln(0.975)$, which is about 1000 years. These are back-of-the-envelope calculations -- any 'real' values used for the calculation could anyway be debated by anybody. Yet, the outcome will always be more or less this order of magnitude. In other words, either the Vostok plots should show a delay between {\COc} and $T$ of the order of 1000 years, or the carbon dioxide is not climate forcing. The data, however, show a delay of $-900$ years\cite{indermuehle} or zero, the latter value resulting from questionable corrections. As far as we know, no correction was proposed to result in the +1000 yr delay necessary to explain contemporary behavior.

What is more, modern correlation figures such as given in Fig.\ \ref{fig:AlGore} also include methane CH$_4$ (available at NOAA Paleoclimatology\cite{NOAA}) and, remarkably, this methane shows the same correlation with {\COc} and $T$. This leaves us flabbergasted. We know that methane is also (assumed to be) a strong climate-forcing greenhouse agent. The enigma is then, How did the information from the {\COc} variations percolate to [CH$_4$] variations? Was this information from {\COc} transmitted to the methane through the temperature variations? In other words, [CH$_4$] variations are the \textit{result} of $T$ variations, rather than their cause? Then we may equally assume that {\COc} variations are the \textit{effect} of $T$ variations rather than their \textit{cause}.

There are several mechanisms that may explain such an inverse phase relation, such as outgassing of {\CO} (and CH$_4$) from the warming oceans and thawing permafrost, the correlation between {\COc} and [CH$_4$] then stems from a common underlying cause. If that is the case, artificially changing the {\CO} in the atmosphere will not change the temperature of our planet, just like heating up a can of soda will liberate the gases contained therein into the atmosphere, while increasing the concentrations of gases above the can of soda will not raise its temperature. This unidirectional relation between temperature and gas concentrations is what is called Henry's Law; the ratio of concentrations of gas dissolved in the liquid and mixed in the air above it in equilibrium is a parameter that depends on temperature. Al-Anezi and coworkers have studied this effect in more detail in a laboratory setup under various conditions of salinity and pressure, etc. For {\CO} in and above water an increase in temperature will cause outgassing with a proportionality that is consistent with the correlation found by the historic correlations of global temperature and {\CO} in the atmosphere. Also, Fischer and coworkers find the delay of {\COc} relative to $T$, as discussed above, likely caused by this ocean outgassing effects\cite{fischer} and find that at colder times, the delay is longer, which is itself consistent with Arrhenius-like behavior of thermally-activated processes, such as most in nature. In the presence of an alternative explanation, there is room for doubt in the AGW ideas that increased {\COc} will cause an increased temperature.

Inspired by this uncertainty in the (Anthropogenic) Global Warming model, we tried to see if we can find more evidence for this failure of the cause-and-effect idea. We looked at the recent historic climate data (from just before the AGW model prevalence) and meticulously-measured {\COc} data and came to the same conclusion, as we will present here.

\section{Results}

We started with the data of a climate report from before the Global Warming claims. We deem these data more reliable since they were for sure not produced under the tutelage of a political committee (IPCC). At least we are more convinced about the neutrality of the scientists reporting these data. Moreover, the work contains all the useful data and are even available on-line; The ideas presented here do not need recent data and thus we refrained from looking at them altogether.

The authors of the work, Balling and coworkers\cite{balling} analyzed the global warming (without capitals because it is not the name of a model) and concluded "Our analysis reveal a statistically significant warming of approximately 0.5 $^{\rm o}$C over the period 1751 to 1995. The period of most rapid warming in Europe occurred between 1890 and 1950, ... no warming was observed in the most recent half century". Note that at the onset of the Global Warming ideas, no warming was observed that can be correlated to the (accelerated) increase of {\COc}. Note also that since 1998 it has not warmed up at all, as confirmed by satellite data (1998 was the warmest year)\cite{NSSTC}, in spite of the continuing exponential increase in atmospheric {\CO}\cite{ESRL}. The temperature seems to be unaffected by the anthropogenic {\CO}.

\begin{figure}
 \centering
 \scalebox{0.55}{\includegraphics{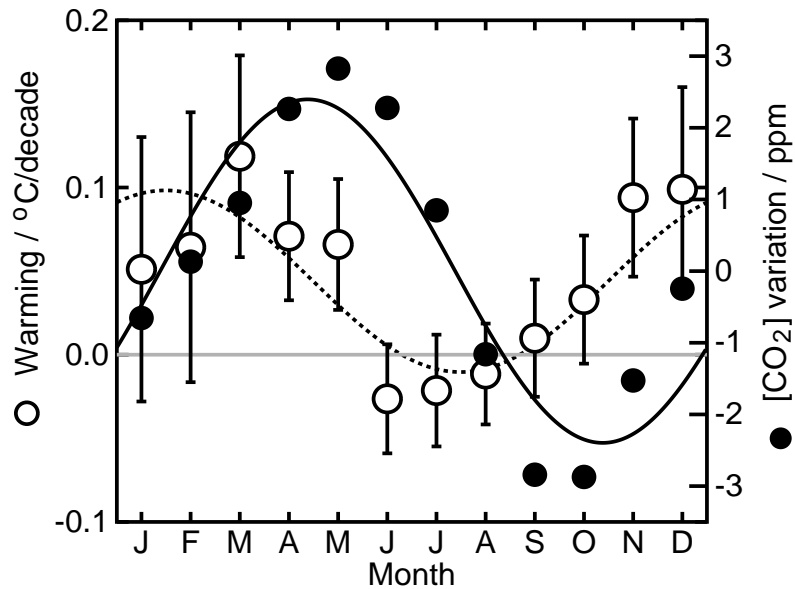}}\\
 \caption{\label{fig:year}
Distribution of global warming (degrees per decade) between 1851 and 1991 (source: Balling, et al.\cite{balling}) and {\CO} concentrations (measured at Mauna Loa, source: NOAA\cite{noaa}) over a year. The dashed and solid lines are sinusoidal fits to the data of temperature and {\COc}, respectively}
\end{figure}

Balling and coworkers then went on to analyze the increase in temperature as a function of the time of the year for the data between 1851 and 1991. They calculated for each of the twelve months the increase in temperature. They found a distribution as given in Figure \ref{fig:year} (open circles).

This figure based on the data of Balling is again remarkable. The first thing we note is that, while there has been an average of warming, this is not spread equally over the year. In fact, summer months have become cooler. Without knowing the underlying reason, this is remarkable, since {\COc} has increased in all months. There are seasonal fluctuations of the {\CO} concentrations, see the black dots which represent the monthly {\COc} fluctuations relative to the yearly average at the Mauna Loa site (source: NOAA, visited 2008\cite{NOAA}). These rapid fluctuations are mainly attributed to biological activity (the Northern hemisphere has more land and in colder times - in winter - more plants are converted into {\CO} and in warmer times - in summer - more photosynthesis takes place converting {\CO} into biomass, i.e., {\COc} is a natural function of temperature). Part of the fluctuations, however, are attributed to human activity (in winter the Northern hemisphere - where more people live - is cold and humans thus burn more fuel to warm their houses, i.e., {\COc} is a function of temperature). As a side note, these two things show us that it is very straightforward to understand how {\COc} can be a function of temperature, in these cases through biological activity, including that of humans, in this case resulting in a rapid inverse proportionality (warmer $\rightarrow$ less {\CO}). Other, long-term processes such as degassing of oceans can have opposite effects, i.e., warmer $\rightarrow$ more {\CO}. While we bear this in mind, we will continue the reasoning of Anthropogenic Global Warming and assume an opposite correlation, that is, temperature is a function of {\COc}, and analyze the oscillations. We will show that this assumption is inconsistent with the data.

While the natural oscillations have always existed and thus don't result in seasonal oscillations of global warming, the human-caused fluctuations should be represented in the temperature fluctuations. What we would expect in the framework of AGW is that all months have warmed up (because of general injection of anthropogenic {\CO} into the atmosphere), but winter months a little bit more (because of seasonal fluctuations of these injections). As a response to the sinusoidal {\COc} oscillations, a sinusoidal oscillation in temperature is to be expected that is i) offset vertically by an amount to make it fully above the zero line, ii) offset (delayed in time) by a time that can be up to 3 months maximum, as will be discussed here. Neither is the case.

\begin{figure}
 \centering
 \scalebox{1}{\includegraphics{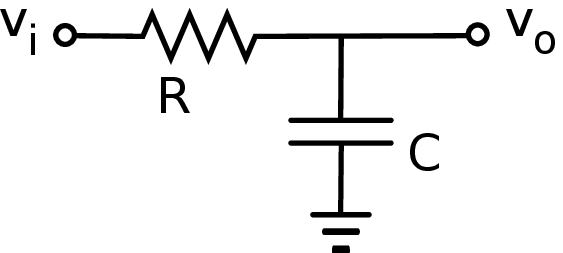}}\\
 \scalebox{0.5}{\includegraphics{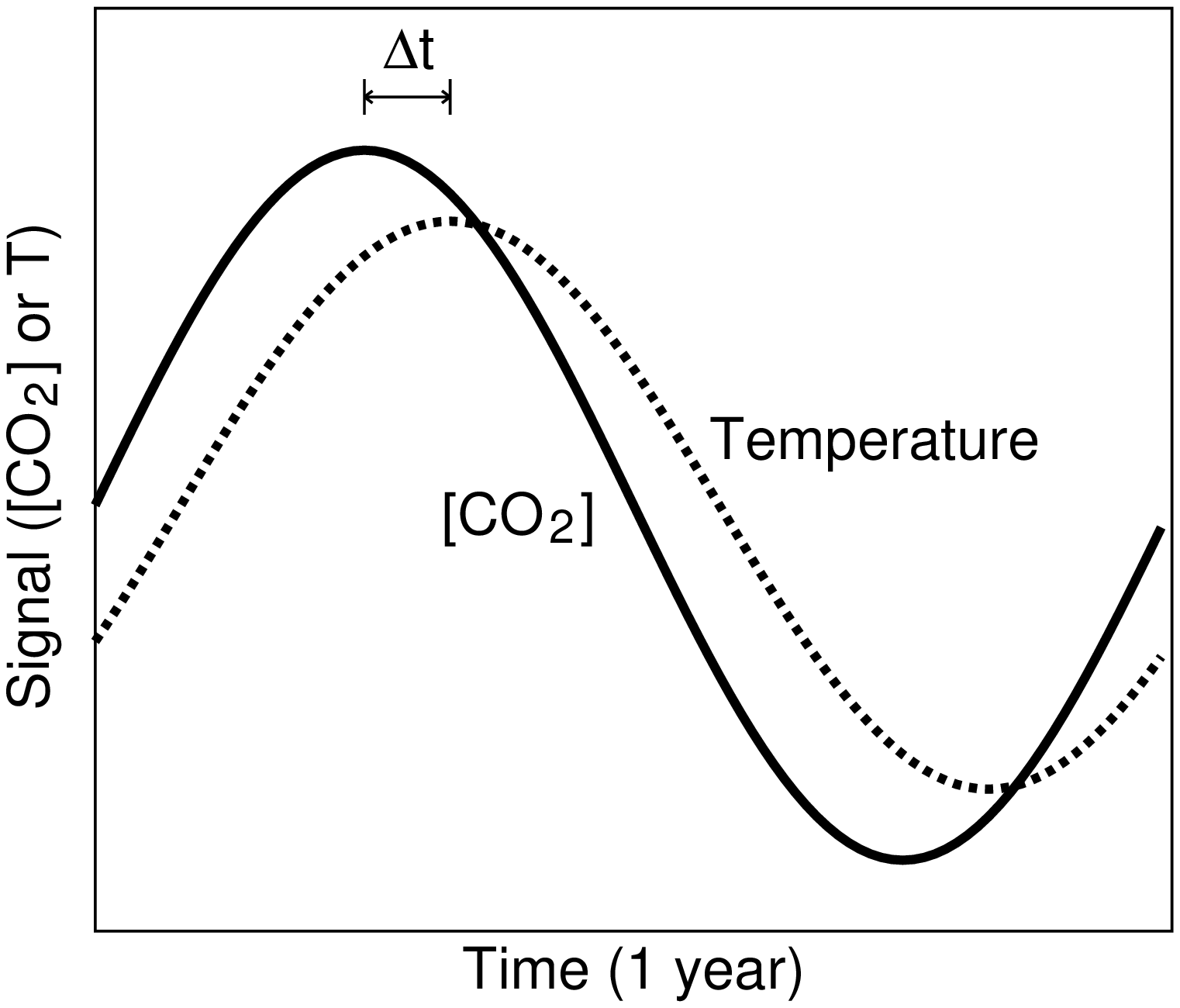}}\\
 \scalebox{0.5}{\includegraphics{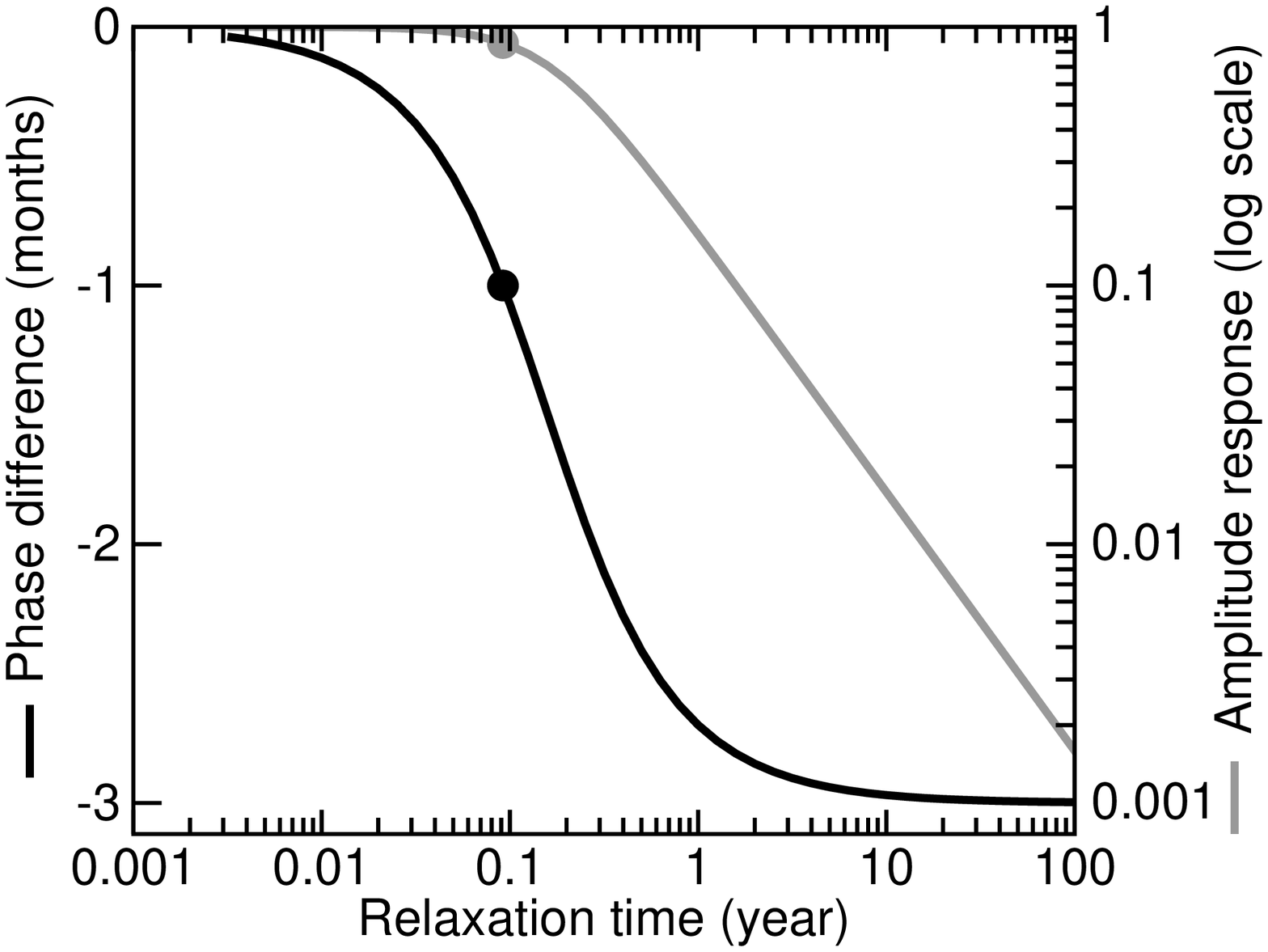}}\\
 \caption{\label{fig:relax}
Relaxation RC system. Top: equivalent circuit. Middle: Temporal response. Bottom: Phase difference (dark curves) and amplitude (gray curve, log scale) of response. Note the maximum phase difference: 3 months. The dot represents the radiation-temperature system, where the warmest (coldest) day is about one month after the 'longest' ('shortest') day, indicating a relaxation time of about 0.1 year (1.2 months)}
\end{figure}

Comparing the monthly fluctuations in temperature increase with monthly fluctuations in {\COc} we see again that the latter lags behind, this time by about 3 months (to be precise, fitting sine curves to the data give a difference of 2.9 months). One might think that the temperature lags behind 9 months -- after all, months are periodic -- but upon second thought, this is not possible. This is best explained in a relaxation model.

Electronic engineers model things with electronic circuits and this case of temperature and {\CO} is also very adequately studied by such circuits. Using an equivalent electronic circuit does not mean that the processes are electronic, but that they can be modeled by such circuits, as in an analog computer. (The appendix gives the mathematical link between a relaxation model and the equivalent electronic circuit).

In this case we have a model between driving force (either {\COc}, as we are wont to believe, or temperature $T$) and the response (respectively, $T$ or {\COc}). For instance, an increase in {\COc} will cause an increase in $T$ by the greenhouse effects. This is necessarily a simple relaxation system, where the changes of the force cause the system to be off-equilibrium until a new equilibrium is reached. This restoring of the equilibrium comes with a certain relaxation time. The reasons for relaxation can be various. For instance, {\CO} has to diffuse to places where it can do its temperature effect. There can even be more than a single relaxation process, and instead be a complicated multi-relaxation process comparable to multi-stage nuclear decay. The fact is that one of the relaxation times is dominant, and we can describe the relaxation by a single relaxation time (that is the sum of all relaxation times). As long as there is no resonance in the system (something that can only be achieved with positive feedback) it will behave as described here.

We will model our climate system with a simple electronic relaxation system consisting of a resistance and a capacitance, $R$ and $C$ respectively, see Figure \ref{fig:relax}. The product of the two yields the relaxation time, $\tau = RC$. At the entrance of this system we connect our  oscillating driving voltage $V_{\rm i}(t)$ (representing, for example, {\COc} oscillations), in which $t$ is time. The response is measured as the charge $Q(t)$ in the capacitor which represents for instance the temperature variations. This charge is also measured by the output voltage by the standard capacitor relation $V = Q/C$. Thus our output voltage $V_{\rm o}(t)$ represents the response (for example temperature).

Applying a sinusoidal input signal, $V_{\rm i}(t) \propto \sin(2\pi f t) \propto$ \COc $(t)$ (with $f$ the frequency of oscillation) we get a sinusoidal wave at the output, with the same frequency, but with a phase at the output that is not equal to the phase at the input signal, $V_{\rm o}(t) \propto \sin(2\pi f t + \theta)$. The phase difference $\theta$ is directly and uniquely determined by the relaxation time of the system $\tau$ and the oscillation frequency $f$, see Figure \ref{fig:relax}.

For very low oscillating frequencies, the system can easily relax and the phase of the output signal is equal to that of the input signal. For increased frequencies or for increased relaxation times the system has difficulty accompanying the driving force. The amplitude at the output drops and starts lagging behind the input. The maximum phase difference for infinite frequencies or infinite relaxation time is exactly one-quarter period.

In our case our oscillating period is one year. One quarter period is thus 3 months and that is the maximum delay we can expect between driving force and response. For relaxation times much longer than the oscillating period of one year, that is the delay one expects. The delay time provides information about the system.

As an example, the comparable system of solar radiation and temperature -- comparable in that the oscillating period is one year and both deal with the weather and climate -- has a delay of one month; the solar radiation and temperature oscillate with one year period, but the warmest day is nearly everywhere one month after the day with the most daylight and the on average coldest day is one month after the day with least daylight. In Figure \ref{fig:relax}(bottom) we see that the relaxation time of the \{radiation $\rightarrow$ temperature\} system therefore must be about 0.1 year (1.2 months). In the plot this is indicated with a dot.

We can get a similar estimation value of the relaxation time of the atmosphere temperature through daily oscillations. As a rough figure, the temperature drops by about 4 degrees at night in about 8 hours after the sun has set. Assume that the relaxation upon this step-like solar radiation is a simple exponential (situation b shown in the appendix) and would finish eventually at close to absolute zero (say 10 kelvin), and starts at 290 K, 4 degrees in 8 hours, we solve the equation
\begin{equation}
(280 \;{\rm K})\times\exp\left(-\frac{(8\;{\rm hours})}{\tau}\right) = (280-4)\;{\rm K},
\end{equation}
which yields 23 days, similar to the value found above from yearly oscillations.

Going back to the data of {\COc} and temperature (Fig.\ \ref{fig:year}) we can now understand the behavior, that is, the phase difference. But only if we assume the temperature to be the driving force. For instance: for some reason the temperature has increased more in winter months, and, as a result, to the natural {\COc} oscillations has been added a component with a maximum in spring months. The alternative, {\COc} being the driving force and a delay of 9 months (3 quarter periods) is mathematically not possible. Another explanation, which we do not consider a valid alternative, the temperature might be lagging behind {\COc} if it has a negative gain, i.e., {\COc} increments lower the temperature. This negative sign of the gain would add another 180$\rm ^o$ phase shift and a total apparent phase shift of 270$\rm ^o$ would be possible. This goes even more against AGW models and we do not see an easy physical explanation how {\CO} might lower the temperature.

\section{Discussion}

This simple analysis opposes the hypothesis that {\COc} is causing serious temperature rises. As said, the model assumes that no resonance occurs that can possibly cause longer delay times. This, in our opinion, is a valid assumption since resonance is not likely. First of all, for this strong positive feedback effects would be needed and they are not likely. Although many climate scientists have proposed positive feedback as discussed in the introduction and they make heavy use of them in order to explain and model the needed non-linear behavior of the greenhouse effect, this goes against intuition. In a chaotic system these feedback factors are then extremely critical. Scientists of any plumage, when making such simulations, know this; if they change their parameters just slightly (sometimes even in the scale of the numerical resolution of their floating point numbers), the outcome can be hugely different.

There is also an experimental argument against positive feedback factors, namely the conscientious satellite measurements, see for instance the work of Lindzen and Choi\cite{lindzen}, Roy Spencer\cite{spencer}, or Wielicki et al.\cite{wielicki}. These, in fact, prove a \textit{negative} feedback in the climate system.
Without feedback, in standard theory, if the Earth warms up (by global warming in a radiation imbalance), the temperature rises and the outward Earth radiation increases by a certain amount, until establishing a new equilibrium. In the AGW model, a positive feedback is used of the form: if the temperature increases, the outward Earth radiation is less than that predicted by standard theory or the incoming solar radiation increases because of reasons like cloud (non)forming, thus increasing the temperature even further. The contrary can also happen: in negative feedback, if the planet heats up by a radiation imbalance for whatever reason, new channels of Earth radiation can be opened or incoming solar radiation blocked (for instance, by increased cloud cover), thus reducing the temperature with respect to standard theory. As demonstrated by the scientists mentioned above, the Earth climate is a negative-feedback 'auto-stabilizing' system, without going into detail what these feedbacks are. This is also in agreement with the fact that, whereas the conditions on our planet have significantly changed over the geological history (the sun for instance has been 25\% less bright than today), the climate has been rather stable, always restoring from climate perturbations to median values instead of saturating in extreme values; the latter one would expect in a thermal-runaway positive-feedback climate system. Note that, if large positive feedback exists, the temperature is unstable and will change until it saturates, that is until negative feedback becomes important. In other words, it is technically not even possible that we are in a positive-feedback situation, considering the stable temperatures. (Compare this to the positive-feedback of a shop-a-holic -- buying always makes him buy even more -- his funds are acceleratingly depleted or his credit increasingly rising, until the banks put a lid on his spending, i.e., negative feedback). We \textit{must} be in a negative-feedback situation and Lindzen and Choi, Spencer, and Wielicki, et al., have proven this by measurements. Negative feedback was already argued to be significant when the consensus of the scientists was for a global cooling, see the work of Idso\cite{idso}.

Additional arguments against positive feedback come from the fact that every day, and every year the temperature system is brought off equilibrium. At night it cools down, in the daytime it warms up. In the winter it cools down and in summer it warms up. These temperature disturbances are much larger and much faster than those that may have been produced by greenhouse gases (20 degrees/day or 30 degrees/year vs 0.7 degrees/100 years). The same accounts for {\CO} disturbances. The human-caused {\CO} is insignificant compared to the large and noisy emissions naturally occurring on this planet (only the accumulated effect of the tiny human-originated {\CO} is supposed to have an effect). To give an idea, Segalstad and coworkers  established that of the current rise in {\COc} levels relative to the preindustrial level, only 12 ppm is attributable to human activity while 68 ppm is attributed to natural phenomena\cite{segalstad}. These fluctuations are also visible in the extensive summary of Beck\cite{beck} and show that even in recent history the {\COc} levels were sometimes higher than the modern values, while as everyone knows, the human emissions have monotonously increased, showing that these huge fluctuations can only have a natural origin. Relevant for the discussion here, the fluctuations would rapidly push the climate off equilibrium if it were unstable.

Yet, in spite of these huge disturbances, both in temperature and {\CO}, the  equilibrium is restored every day and every year and every century. Had the earth climate been a positive-feedback system, in summer or in winter the temperature would have been in a runaway situation, unrecoverable in the following compensating half-period. Apparently the system can recover very easily and repeatedly from such huge disturbances. The reason is that the climate is a negative-feedback system that stabilizes itself. This is an unavoidable conclusion.

One might think that the seasonal fluctuations are too fast to be causing a runaway scenario and that before the system runs away it already recovers. That is a misapprehension; changes cannot be too fast. If the system is unstable, it is unstable. If starting oscillations are much faster than the response time of the system, the effective amplitude is reduced, but in a runaway system they will be amplified up to the point of saturation. The system can only be stable if the feedback factor at that specific frequency is not positive. Look at it like this: In the first half of the year, it is hot and the system tries to runaway. In the second half of the year it is colder and it will restore, but it has a minute memory that the temperature has already run off a little in the first half and the second half therefore does not compensate completely. In the first year we remain with a tiny temperature offset. Once this offset is introduced, the system will runaway. Of course, it can runway in both directions. Chance will determine which one, but if the system is unstable (positive feedback), the system will runaway. Like the metastable system of a ball placed on top of a hill. It can only stay there in the absence of noise or any fluctuation in general. In conclusion, only negative feedback makes sense.

Relevant to the current work, such negative feedback will make any delay longer than 1/4 period impossible. Thus, the fact that we find a delay close to a quarter period means that i) The temperature signal is the origin for {\COc} signal (or the two are uncorrelated) and ii) the relaxation time $\tau$ linking the two is (much) longer than the period (12 months) of oscillation.

Moreover, even if positive feedback were present, for the resonance itself to be significant, the oscillating frequency needs to be close to the resonance frequency, i.e., 12 months. It is highly unlikely that the natural frequency of the climate-{\COc} system is close to the 12-months-periodic driving force. Even more so, since also the long-term ice-drilling data need to be explained somehow, where delays of several thousands of years are observed. In our analysis, relaxation times of several thousands of years will explain both the ice-drilling data, as well as the yearly temperature and {\COc} oscillations.

Finally, the set of data we used is rather limited. We only used data presented by Balling, et al., that ends at the end of the 20th century. Moreover, they only have data from the Northern Hemisphere. Future research should tell if the ideas presented here can stand up to scrutiny when more recent data and pan-global data are used. As a note of proof, Humlum et al.\cite{humlum}, have recently investigated correlation between temperature and {\COc} variations on the time scale of decades, similarly concluding that {\COc} changes are delayed in relation to temperature, and can therefore not be the reason for temperature changes.

\section{Conclusion}

In conclusion, the idea tested here that {\COc} is the \textit{cause} of temperature changes does not pass our signal analysis. It goes a little too far to say that this what we present here is proof for the opposite, namely that {\COc} is the \textit{effect} of temperature, but our analysis does not contradict this. Future will tell if such an hypothesis may be postulated with some confidence.

Acknowledgements: This research was paid by no grant. It received no funding whatsoever, apart from our salaries at the university where we work. Nor are we members of any climate committees (political or other) or are we linked to companies or NGOs, financially or otherwise. This is an independent opinion that does not necessarily represent the opinion of our university or of our government.


\section{Appendix: The mathematics of relaxations}

In simple relaxation models the (negative) change of a quantity is proportional to the magnitude of the remaining quantity. Simple examples are nuclear decay, in which the change of number $N$ of atoms at a certain time $t$ is given by ${\rm d}N(t)/{\rm d}t = -\alpha N(t)$, or the velocity $v$ of an object under friction is given by ${\rm d}v(t)/{\rm d}t = -\beta v(t)$. ($\alpha$ and $\beta$ positive constants). From experience, and by solving the differential equation, we know that such systems show exponential decay, $N(t) = N_0\exp(-\alpha t)$ and $v(t) = v_0\exp(-\beta t)$ respectively.

\begin{figure}
 \centering
 \scalebox{1}{\includegraphics{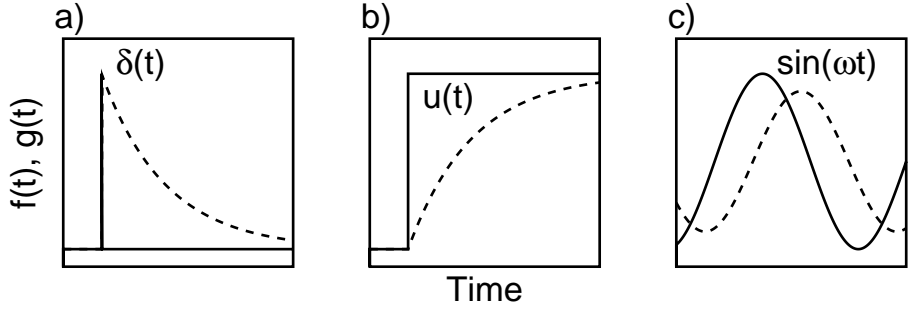}}\\
 \caption{\label{fig:apndx}
Cause and effect functions, $f(t)$, solid lines, and $g(t)$, dashed lines, for relaxation systems, with three different driving-force functions: a) $f(t) = \delta(t)$, b) $f(t) = u(t)$, c) $f(t) = \sin(\omega t)$}
\end{figure}

Now, we can take a function $f(t)$ that is the driving force of another quantity $g(t)$, the response function, respectively the cause and the effect. We can decompose the function $f$ into an integral of Dirac-delta functions. The response to each delta function is given by the function $d(t)$. Assuming linearity, the total response is then a convolution
\begin{eqnarray}
\nonumber
g(t) &=& \int_{-\infty}^\infty d(s)u(s)f(t-s){\rm d}s\\
&=& \int_0^\infty d(s)f(t-s){\rm d}s,
\end{eqnarray}
where the Heaviside function $u(s)$ ($u(s)=1$ for $s>0$ and 0 otherwise) was used to force the causality; the response $d(s)$ can only come after the driving force. (Note that non-linearities will not change the sign of these calculations, i.e., a delay cannot become an advance.) For instance, if the response function is an exponential decay, as mentioned above,
\begin{equation}
\label{eq:conv}
g(t) = \int_0^\infty g_0\exp(-\alpha s)f(t-s){\rm d}s
\end{equation}
Substituting a delta-function at $t=0$ for the driving force $f$ will reproduce the exponential decay:
\begin{eqnarray}
\nonumber
g(t) &=& \int_0^\infty g_0\exp(-\alpha s)\delta(t-s){\rm d}s\\
\nonumber
&=& \int_0^\infty g_0\exp(-\alpha [t-s])\delta(s){\rm d}s\\
&=& g_0\exp(-\alpha t)u(t)
\end{eqnarray}
In other words, the response to a 'spike', a delta function at $t=0$ is an exponential decay with an amplitude $g_0$, and time constant $\tau=1/\alpha$. The response to a Heaviside (step)function $f(t) = u(t)$ is then given by
\begin{eqnarray}
\nonumber
g(t) &=& \int_0^\infty g_0\exp(-\alpha s)u(t-s){\rm d}s\\
\nonumber
&=& \int_0^t g_0\exp(-\alpha s){\rm d}s\\
&=& g_0u(t)[1-\exp(-\alpha t)]/\alpha
\end{eqnarray}
More interesting -- more relevant for our work -- is the case of a sinusoidal driving force. This can now easily be calculated by substituting the driving-force function $f$ into Eq.\ \ref{eq:conv}:
\begin{eqnarray}
f(t) &=& f_0\sin(\omega t)\\
\nonumber
\label{eq:grad}
g(t) &=& \int_0^{\infty} g_0 \exp(-\alpha s)f_0\sin[\omega (t-s)]{\rm d}s\\
\nonumber
&=& \frac{f_0g_0}{\alpha^2+\omega^2} \left[ \alpha\sin(\omega t) - \omega\cos(\omega t)\right]\\
&=& \frac{f_0g_0}{\sqrt{\alpha^2+\omega^2}} \sin\left(\omega t - \tan^{-1}[\omega/\alpha]\right)
\end{eqnarray}
(For the second step in Eq.\ \ref{eq:grad} Gradshteyn and Ryshik\cite{gradshteyn} was used). Figure \ref{fig:apndx} shows these three cases of driving forces and response functions. Figure \ref{fig:AlGore} shows a simulation with the driving function $f(t)$ equal to the measured temperature and a delay of $\tau$ ($= 1/\alpha$) = 15 kyr, which results in a quite good representation of the \COc curve.

An electronic circuit such as presented here has these properties of exponential response to a Heaviside function and linearity and the response of Equation \ref{eq:grad}. For this reason, such (virtual) circuits are widely used in simulations of phenomena including phenomena far away from electronics.
The interesting and relevant conclusion of Eq.\ \ref{eq:grad} is that the maximum phase shift is 90$^{\rm o}$ and this occurs for frequencies that are much higher than the relaxation speed, $\omega \gg \alpha$.

\end{document}